\documentstyle[12pt]{article}

\newcommand{\be}{\begin{equation}}
\newcommand{\ee}{\end{equation}}
\newcommand{\ba}{\begin{eqnarray}}
\newcommand{\ea}{\end{eqnarray}}

\newcommand{\no}{\nonumber\\}

\newcommand{\grts}{\raise.3ex\hbox{$>$\kern-.75em\lower1ex\hbox{$\sim$}}}
\newcommand{\lets}{\raise.3ex\hbox{$<$\kern-.75em\lower1ex\hbox{$\sim$}}}

\title{\bf Soft CP breaking \\ and the strong-CP problem}

\author{L.\ Lavoura \\
\small Universidade T\'ecnica de Lisboa \\
\small CFIF, Instituto Superior T\'ecnico,
Edif\'\i cio Ci\^encia (f\'\i sica) \\
\small P-1096 Lisboa Codex, Portugal}

\begin{document}

\maketitle

\begin{abstract}
I put forward a model with vectorlike isosinglet quarks
in which soft breaking of CP
leads to strong CP violation only arising at two-loop level.
\end{abstract}

Non-perturbative effects in Quantum Chromodynamics (QCD)
may lead to P and CP violation,
characterized by a parameter $\theta$,
in strong interactions.
The experimental upper bound on
the electric dipole moment of the neutron
requires $\theta \lets 2 \times 10^{-10}$ \cite{baluni}.
(Electric dipole moments of elementary particles
are both P- and T-violating;
in normal field theory,
T violation is equivalent to CP violation.)
The presence of this unnaturally small number in QCD
is known as the strong-CP problem.

$\theta$ is the sum of two terms,
$\theta_{QCD}$ and $\theta_{QFD}$.
$\theta_{QCD}$ is the original value of the angle $\theta$
characterizing the QCD vacuum.
$\theta_{QFD}$ originates in the chiral rotation of the quark fields
needed to render the quark masses real and positive.
If $M_p$ and $M_n$ are the mass matrices of the up-type (charge $2/3$)
and down-type (charge $-1/3$) quarks,
then $\theta_{QFD} = \arg \det (M_p M_n)$.

There are two general ways of solving the strong-CP problem.
In the first approach
$\theta$ is claimed to have no significance or physical consequences;
theories with different values of $\theta$ are equivalent,
and we may set $\theta = 0$ without loss of generality.
This may be so because of the presence in the theory
of a Peccei--Quinn symmetry \cite{peccei},
but there are also claims
that QCD dynamics itself cures the strong-CP problem \cite{emilio}.
The second path,
which I shall follow,
tries to find some symmetry
which naturally leads to the smallness of $\theta$.
As $\theta$ is the coefficient of a quartic term in the Lagrangian
which is both CP- and P-violating,
one first assumes the Lagrangian
--- or at least its quartic part ---
to be either CP- or P-symmetric,
thereby automatically obtaining $\theta_{QCD} = 0$.
CP or P symmetry must then be either softly or spontaneously broken.
While doing this
the problem of ensuring the smallness of $\theta_{QFD}$ remains.
We must obtain quark mass matrices which,
even though they are complex
--- because after all
we know that CP is violated in the electroweak interactions ---
are such that $\det (M_p M_n)$ is real.
Afterwards,
we must still ensure that the quantum (loop) corrections
to the mass matrices
do not lead to a much too large $\theta_{QFD}$.

In this letter I put forward a model
in which the quartic terms of the Lagrangian are assumed
to enjoy CP symmetry,
thereby setting $\theta_{QCD} = 0$.
CP symmetry is softly broken by the mass terms,
which have dimension three,
of isosinglet vectorlike quarks.
The assumption that a symmetry is broken only by terms of cubic dimension
is technically natural,
in the sense that it is not corrected by infinite renormalizations.
Therefore,
there will be no non-real divergent renormalization
of the quartic terms in the Lagrangian.
Because of a particular discrete symmetry,
$\theta_{QFD}$ vanishes at tree level.
I explicitly show that $\theta_{QFD}$ also vanishes at one-loop level.
My model has some similarity with two previous ones \cite{babu,bento},
but is simpler and more economic than any of them.

The gauge group is SU(2)$\otimes$U(1) just as in the standard model (SM).
The scalar sector consists of the usual doublet $\phi$ of the SM
(with $\tilde\phi \equiv i \tau_2 \phi^\ast$)
together with a real singlet $S$.
The lepton sector is as in the SM.
The quark sector
consists of the usual three left-handed doublets
$q_L = (p_L , n_L)^T$,
the three right-handed up-type quarks $p_R$,
and the three right-handed down-type quarks $n_R$.
Besides,
I also assume the existence
of $l$ SU(2)-singlet vectorlike quarks of charge $-1/3$,
with left-handed components $N_L$ and right-handed components $N_R$.
The number $l$ is arbitrary.

The model has a discrete symmetry
under which both $S$ and the $N_R$ change sign
while all other fields remain unaffected.
As a consequence,
the scalar potential is
\be \label{potential}
V =
\mu_1 \phi^\dagger \phi
+ \lambda_1 (\phi^\dagger \phi)^2
+ \mu_2 S^2
+ \lambda_2 S^4
+ \lambda_3 S^2 \phi^\dagger \phi
\ee
and the Yukawa Lagrangian is
\be \label{Yukawa}
\cal{L}_Y =
- \overline{q_L} \phi \Gamma n_R
- \overline{q_L} \tilde\phi \Delta p_R
- \overline{N_L} S \Upsilon N_R + H.c.,
\ee
where $\Gamma$ and $\Delta$ are $3 \times 3$ matrices,
$\Upsilon$ is an $l \times l$ matrix.
The mass terms $- \overline{N_L} M n_R + H.c.$,
where $M$ is an $l \times 3$ matrix,
are relevant too.
The vacuum expectation value (vev) of $\phi$ is $v e^{i \alpha}$
and the one of $S$ is $V$.
I decompose the scalar multiplets as
\be \label{scalardecomposition}
\phi = e^{i \alpha}
\left(\!
\begin{array}{c}
\varphi^+ \\
v + (H + i \chi) / \sqrt{2}
\end{array}
\!\right),\ \
S = V + R.
\ee
$\varphi^\pm$ and $\chi$ are the charged and neutral Goldstone bosons,
to be absorbed as the longitudinal components
of the $W^\pm$ and $Z$ gauge bosons,
respectively.
After using the vacuum-stability conditions
to eliminate $\mu_1$ and $\mu_2$ in favor of $v$ and $V$,
the Higgs potential in Eq.~(\ref{potential})
yields the mass terms for $H$ and $R$:
\be \label{developedpotential}
V = 2 \lambda_1 v^2 H^2
+ 4 \lambda_2 V^2 R^2
+ 2 \sqrt{2} \lambda_3 v V H R
+ ...
\ee
It follows that $H$ and $R$ mix to form the two physical scalar particles,
which I shall denote $S_1$ and $S_2$.
I denote the mixing angle $\psi$.

At tree level
\be \label{treelevel}
\theta_{QFD} = \arg \det
[ (e^{i \alpha} v \Gamma) (e^{- i \alpha} v \Delta) (V \Upsilon) ]
= \arg \det (\Gamma \Delta \Upsilon)
\ee
does not depend on $\alpha$.

Let us denote
$M_u = {\mbox{diag}} (m_u, m_c, m_t)$
the diagonal matrix of the masses of the three up-type quarks,
and $M_d = {\mbox{diag}} (m_d, m_s, m_b, ...)$
the diagonal matrix of the masses of the $l+3$ down-type quarks.
Without loss of generality,
we may choose to work in the weak basis in which
$p_L = u_L$ and $p_R = u_R$ are the physical up-type-quark fields,
and $\Delta = M_u / (v e^{- i \alpha})$.
In this basis,
we must bi-diagonalize the mass matrix of the down-type quarks.
This we do by performing the unitary transformations
($A$ may be $L$ or $R$)
\be \label{transformation}
\left( \begin{array}{c}
n_A \\ N_A
\end{array} \right)
= \left( \begin{array}{c}
X_A \\ Y_A
\end{array} \right)
d_A,
\ee
in which $X_A$ is a $3 \times (l+3)$ and $Y_A$ is an $l \times (l+3)$ matrix.
Indeed,
$X_L$ is the Cabibbo--Kobayashi--Maskawa (CKM)
mixing matrix of the charged gauge interactions.
I define the $(l+3) \times (l+3)$ hermitian matrices
$H_A \equiv X_A^\dagger X_A$.
As a consequence of the unitarity
of the transformation matrices in Eq.~(\ref{transformation}),
$Y_A^\dagger Y_A = 1 - H_A$,
and $H_A^2 = H_A$.
The matrix $H_L$ appears in the neutral-current interactions
of the down-type quarks \cite{branco},
which are given by
\be \label{Zinteractions}
\frac{g}{\cos \theta_W} Z_\mu
\overline{d} \gamma^\mu
\left( \frac{\sin^2 \theta_W}{3} - \frac{1 - \gamma_5}{4} H_L \right)
d.
\ee
The decays of the $Z$ measured at LEP,
together with the observed suppression
of neutral flavor-changing interactions
and the near unitarity of the CKM matrix,
impose strong constraints on the $3 \times 3$ part of $X_L$
corresponding to the three known charge $-1/3$ quarks \cite{silva}.
I shall assume that both $M$ and the vev $V$ are very large
--- in the case of $M$ because it consists of singlet mass terms,
in the case of $V$ because it is the vev of a singlet field,
and it does not break the electroweak gauge group.
Under that assumption,
the $3 \times 3$ part of $X_L$ is almost unitary,
as required.
Moreover,
the mixing angle $\psi$ between $H$ and $R$ scales as $v/V$
[see Eq.~(\ref{developedpotential})].
Therefore,
in the limit of large $V$,
$R$ is an eigenstate of mass with mass $\sim V$.

The bi-diagonalization condition is
\be \label{diagonalization}
\left( \begin{array}{cc}
v e^{i \alpha} \Gamma & 0 \\ M & V \Upsilon
\end{array} \right)
= \left( \begin{array}{c}
X_L \\ Y_L
\end{array} \right)
M_d
\left(\! \begin{array}{cc}
X_R^\dagger & Y_R^\dagger\!
\end{array} \right).
\ee
It follows that
\ba \label{constraints}
H_L M_d (1 - H_R) & = & 0,
\no
H_R M_d^{-1} (1 - H_L) & = & 0.
\ea
The Yukawa interactions of the fields $H$ and $R$
with the down-type quarks are given by
\ba \label{developedYukawa}
- e^{i \alpha} \frac{H}{\sqrt{2}} (\overline{n_L} \Gamma n_R + H.c.)
& = &
- \frac{H}{\sqrt{2} v} (\overline{d_L} H_L M_d d_R + H.c.),
\no
- R (\overline{N_L} \Upsilon N_R + H.c.) & = &
- \frac{R}{V} [\overline{d_L} M_d (1 - H_R) d_R + H.c.],
\ea
where I have used Eqs.~(\ref{constraints}).

Up to this point
I have not postulated anything about the nature of CP violation
in this model.
I now make the further assumption that
the quartic part of the Lagrangian conserves CP.
Then,
$\theta_{QCD}$ vanishes,
the matrices $\Delta$,
$\Gamma$ and $\Upsilon$ are real,
and $\theta_{QFD} = 0$ at tree level.
CP cannot be spontaneously broken in this model,
because there is no gauge-invariant phase
between the vevs of the scalars
(the phase $\alpha$ can be gauged away).
Therefore,
CP must be softly broken by the only non-quartic terms in the Lagrangian
which may be complex:
the matrix $M$.
As CP-violating amplitudes are $\sim 10^{3}$ times smaller
than CP-conserving ones,
one may naturally assume that the complex phases
in the matrix elements of $M$
are all $\sim 10^{-3}$.

The first model
in which soft breaking of CP was used to suppress strong CP violation
was suggested by Georgi \cite{georgi}.
My present model has a clear similarity
with the one of Bento,
Branco,
and Parada (BBP) \cite{bento}.
Both models are of the Barr--Nelson \cite{nelson} type.
In the model of BBP CP breaking is spontaneous
and there is a complex,
instead of real,
singlet.
There are three neutral scalars,
and there is CP violation via scalar-pseudoscalar mixing,
which leads to strong CP violation arising at one-loop level.
In contrast,
in my model,
just as in the one of Babu and Mohapatra (BM) \cite{babu},
there are only two neutral scalars,
and there is no scalar-pseudoscalar mixing,
allowing strong CP violation to arise only at two-loop level,
as I shall now show.
The proof is analogous to the one valid in the model of BM.

Strong CP violation might arise at one-loop level
via the self-energies of the down-type quarks,
which give rise to complex mass terms.
The relevant diagram is the one in Fig.~1.
Let us denote the mass term originating from that diagram
$\Sigma_d^{ab}$.
$\Sigma_d$ is an $(l+3) \times (l+3)$ complex matrix.
The resulting contribution to $\theta_{QFD}$
is ${\mbox{Im tr}} (M_d^{-1} \Sigma_d)$ \cite{ellis}.
As $M_d$ is diagonal,
only the diagonal matrix elements of $\Sigma_d$ are relevant.
Now,
from Eqs.~(\ref{developedYukawa})
\ba \label{Sigmad}
\Sigma_d^{ab} & = & \sum_{k=1}^{l+3} m_k
\left\{
x (H_L M_d)^{ak} (H_L M_d)^{kb}
+ y [ M_d (1 - H_R)]^{ak} [ M_d (1 - H_R)]^{kb}
\right.
\no
              &   &
\left.
+ z (H_L M_d)^{ak} [ M_d (1 - H_R)]^{kb} 
+ z [ M_d (1 - H_R)]^{ak} (H_L M_D)^{kb} 
\right\},
\ea
where the coefficients $x$,
$y$,
and $z$ are real functions,
resulting from the loop integration,
of the squared mass of the $k^{\mbox{\scriptsize th}}$ down-type quark,
$m_k^2$,
of the masses of $S_1$ and $S_2$,
and of the mixing angle $\psi$.
Then,
using Eqs.~(\ref{constraints}),
one finds
\be \label{trace}
{\rm tr} (M_d^{-1} \Sigma_d) = \sum_{k=1}^{l+3} m_k^2
\left[ (x + z) H_L - (y + z) H_R + y \right]^{kk}.
\ee
As $H_L$ and $H_R$ are hermitian matrices,
their diagonal elements are real,
and ${\mbox{Im tr}} (M_d^{-1} \Sigma_d)$ vanishes.

$\theta_{QFD}$ therefore arises only at two-loop level.
Its magnitude should be suppressed by at least two factors:
a factor $(16 \pi^2)^{-2}$ from the loop integrations,
and a factor $\sim 10^{-3}$ from the small CP-violating phases in $M$.
Other suppression factors of the form $m_q / m_W$,
in which $m_q$ is the mass of either the charm or the strange quark,
may be present too \cite{ellis}.
It should however be noted that in this model,
such as in the one of BBP,
due to the non-diagonal character of $H_L$,
there are diagrams contributing to $\theta_{QFD}$ at two-loop level
with one $Z$ loop and one $W^\pm$ loop,
or with two $Z$ loops.
Those diagrams evade the suppressions derived,
in the context of the SM,
in Ref.~\cite{ellis}.
They should however be suppressed
by the smallness of the non-diagonal matrix elements of $H_L$.
Also,
there are diagrams with scalar loops
contributing to $\theta_{QFD}$ at two-loop level,
involving in particular the mixing matrix $H_R$
[see Eq.~(\ref{developedYukawa})].
Those diagrams will be suppressed if $R$ is very heavy
(if $V$ is very large),
as I advocate.
Summing up,
it is not clear how large may the two-loop contribution
to $\theta_{QFD}$ be in this model.
It is evident anyway that,
by avoiding scalar-pseudoscalar mixing,
my model avoids an important source of strong CP violation present
in the model of BBP,
and in many other models.

I thank Lincoln Wolfenstein for having read a draft of the manuscript
and commented on it.

\end{document}